\documentclass[journal]{IEEEtran}
\usepackage{cite}
\makeatletter
\newcommand\notsotiny{\@setfontsize\notsotiny{6.5}{6.5}}
\makeatother
\usepackage{amsmath,amssymb,amsfonts}
\usepackage{algorithmic}
\usepackage{graphicx}
\usepackage{textcomp}
\usepackage{nomencl}
\usepackage{etoolbox}
\usepackage{hyperref}
\makenomenclature

\begin{document}
\title{Experimental validation of a three-dimensional heat transfer model within the scala tympani with application to magnetic cochlear implant surgery}
\author{Fateme Esmailie, Mathieu Francoeur, Tim Ameel
\thanks{Submission date: 09/01/2020 }
\thanks{Research reported in this publication was supported by the National Institute on Deafness and Other Communication Disorders of the National Institutes of Health under Award Number R01DC013168. }
\thanks{F. Esmailie is with Department of Mechanical Engineering, University of Utah, Salt Lake City, Utah 84112, USA. }
\thanks{M. Francoeur is with Department of Mechanical Engineering, University of Utah, Salt Lake City, Utah 84112, USA .}
\thanks{T. Ameel is with Department of Mechanical Engineering, University of Utah, Salt Lake City, Utah 84112, USA (e-mail: ameel@mech.utah.edu).}}

\maketitle

\begin{abstract}
Magnetic guidance of cochlear implants is a promising technique to reduce the risk of physical trauma during surgery. In this approach, a magnet attached to the tip of the implant electrode array is guided within the scala tympani using a magnetic field. After surgery, the magnet must be detached from the implant electrode array via localized heating and removed from the scala tympani which may cause thermal trauma. Objectives: The objective of this work is to experimentally validate a three-dimensional (3D) heat transfer model of the scala tympani which will enable accurate predictions of the maximum safe input power to avoid localized hyperthermia when detaching the magnet from the implant electrode array. Methods: Experiments are designed using a rigorous scale analysis and performed by measuring transient temperatures in a 3D-printed scala tympani phantom subjected to a sudden change in its thermal environment and localized heating via a small heat source. Results: The measured and predicted temperatures are in good agreement with an error less than 6$\%$. Conclusions: The validated 3D heat transfer model of the scala tympani is finally applied to evaluate the maximum safe input power to avoid localized hyperthermia when detaching the magnet. For the most conservative case where all boundaries except the insertion opening are adiabatic, the power required to release the magnet attached to the implant electrode array by 1 mm$^3$ of paraffin is approximately half of the predicted maximum safe input power. Significance: This work will enable the design of a thermally safe magnetic cochlear implant surgery procedure.  
\end{abstract}


\renewcommand\nomgroup[1]{%
  \item[\bfseries
  \ifstrequal{#1}{A}{Parameters}{%
  \ifstrequal{#1}{B}{Greek symbols}{%
  \ifstrequal{#1}{C}{Subscripts}{}}}%
]}
\mbox{}

\nomenclature[A, 01]{$B$}{Thermal dose constant $\left[-\right]$}
\nomenclature[A, 02]{CEM}{Cumulative Equivalent Minutes $\left[\rm{s}\right]$}
\nomenclature[A, 03]{$c_p$}{Specific heat $\left[\frac{\rm{J}}{\rm{kg} \cdot \rm{K}}\right]$}
\nomenclature[A, 04]{$Fo$}{Fourier Number $\left[-\right]$}
\nomenclature[A, 05]{$g$}{Gravitational acceleration $\left[\frac{\rm{m}}{\rm{s}^2}\right]$}
\nomenclature[A, 06]{$Gr$}{Grashof Number $\left[-\right]$}
\nomenclature[A, 07]{$h$}{Heat transfer coefficient $\left[\frac{\rm{W}}{\rm{m^2 \cdot K}}\right]$}
\nomenclature[A, 08]{${I}$}{Electrical current $\left[\rm{A}\right]$}
\nomenclature[A, 09]{$\mathbf{I}$}{$3\times3$ identity matrix $\left[-\right]$}
\nomenclature[A, 10]{$k$}{Thermal conductivity $\left[\frac{\rm{W}}{\rm{m \cdot K}}\right]$}
\nomenclature[A, 11]{$L_c$}{Characteristic length $\left[\rm{m}\right]$}
\nomenclature[A, 12]{$Nu$}{Nusselt number $\left[-\right]$}
\nomenclature[A, 13]{$P$}{Power $\left[W\right]$}
\nomenclature[A, 14]{$Pr$}{Prandtl number $\left[-\right]$}
\nomenclature[A, 15]{$q$}{Power density $\left[\frac{\rm{W}}{\rm{m^3}}\right]$}
\nomenclature[A, 16]{$\mathbf{{q^{\prime\prime}}}_r$}{Radiative flux $\left[\frac{\rm{W}}{\rm{m^2}}\right]$}
\nomenclature[A, 17]{$Q$}{Dimensionless power $\left[-\right]$}
\nomenclature[A, 18]{$R$}{Electrical resistance $\left[\Omega\right]$}
\nomenclature[A, 19]{$Ra$}{Rayleigh number $\left[-\right]$}
\nomenclature[A, 20]{$t$}{Time $\left[\rm{s}\right]$}
\nomenclature[A, 21]{$T$}{Temperature $\left[\rm{K}\right]$}
\nomenclature[A, 22]{$\mathbf{u}$}{Velocity vector $\left[\frac{\rm{m}}{\rm{s}}\right]$}
\nomenclature[A, 22]{$u$}{Velocity $\left[\frac{\rm{m}}{\rm{s}}\right]$}
\nomenclature[A, 23]{$U$}{Dimensionless velocity $\left[-\right]$}
\nomenclature[A, 24]{$V$}{Volume $\left[\rm{m^3}\right]$}
\nomenclature[A, 25]{$x$, $y$, $z$}{Cartesian coordinates $\left[\rm{m}\right]$}
\nomenclature[A, 26]{$X$, $Y$, $Z$}{Dimensionless length along Cartesian coordinates $\left[-\right]$}
\nomenclature[B, 01]{$\alpha$}{Thermal diffusivity $\left[\rm{\frac{m^2}{s}}\right]$}
\nomenclature[B, 02]{$\beta$}{Thermal expansion coefficient $\left[\rm{\frac{1}{K}}\right]$}
\nomenclature[B, 03]{$\theta$}{Dimensionless temperature $\left[-\right]$}
\nomenclature[B, 04]{$\mu$}{Dynamic viscosity $\left[\frac{\rm{N \cdot s}}{\rm{m^2}}\right]$}
\nomenclature[B, 04]{$\nu$}{Kinematic viscosity $\left[\frac{\rm{m^2}}{\rm{ s}}\right]$}
\nomenclature[B, 05]{$\rho$}{Density $\left[\frac{\rm{kg}}{\rm{m^3}}\right]$}
\nomenclature[B, 06]{$\omega$}{Blood perfusion rate $\left[\frac{\rm{1}}{\rm{s}}\right]$}
\nomenclature[C]{$\rm{bl}$}{Blood}
\nomenclature[C, 01]{$fluid$}{Fluid}
\nomenclature[C,02]{$i$}{Initial}
\nomenclature[C, 03]{$max$}{Maximum}
\nomenclature[C, 04]{$met$}{Metabolic}
\nomenclature[C, 05]{n}{${n^{\rm{th}}}$ time interval}
\nomenclature[C, 06]{$s$}{Surface}
\nomenclature[C, 07]{$0$}{Reference}
\nomenclature[C, 08]{$\infty$}{Ambient}
\nomenclature[C, 09]{$x$}{$x$ component}
\nomenclature[C, 10]{$y$}{$y$ component}
\nomenclature[C, 11]{$z$}{$z$ component}

\printnomenclature

\section{Introduction}
\label{sec:introduction}

Magnetic guidance of cochlear implants is a promising surgical technique that is expected to mitigate physical trauma arising with manual insertion of implants \cite{b48}--\cite{b50}. In this technique, a magnetic field guides the implant electrode array within the scala tympani, one of the cochlear canals, using a small magnet attached to the implant tip. After surgery, the magnet must be detached from the implant electrode array and removed from the scala tympani, since the presence of a magnet in the cochlea may cause medical complications if the patient is exposed to a strong external magnetic such as in magnetic resonance imaging. The removal can be accomplished by melting the substance bonding the implant electrode array and the magnet via localized heating. Consequently, the magnet detachment process may cause localized hyperthermia within the scala tympani, which is one of the more serious impediments preventing the establishment of magnetic cochlear implant surgery \cite{b22}, \cite{b27}. Localized hyperthermia in the ear, which is not unique to magnetic cochlear implant surgery, arises in a variety of applications such as infrared neural stimulation implants \cite{b1} -- \cite{b25}, stapedectomy\cite{b43}, \cite{b44}, caloric test\cite{b41}, \cite{b42}, and radio-frequency radiation devices such as cellular phones\cite{b45} --\cite{b47}. In addition, studies have shown that localized hypothermia can preserve tissues after cochlear implantation \cite{b4, b5}. 

Despite the importance of localized hyperthermia and hypothermia in hearing research, validated three-dimensional (3D) heat transfer computational models within cochlear canals are scarce. Thompson et al. \cite{b1} --\cite{b21} and Liljemalm \cite{b37} developed a heat transfer model to evaluate the thermal impact of infrared neural stimulation implants. In this technique, the implant focuses infrared radiation for stimulating modulus nerves, which are a bundle of nerves located in the middle of the bony cochlea. They validated their numerical results using the experimental data of Shapiro et al. \cite{b3}, who conducted in-vitro experiments by stimulating an “oocyte”. Thompson et al. \cite{b1} --\cite{b21} and Liljemalm \cite{b37} analyzed the feasibility of infrared neural stimulation implants, but the actual geometry of the cochlea was not considered in the validation phase of these works. Similar to magnetic cochlear implant surgery, infrared neural stimulation implant surgery involves localized heating. However, neither the heat source nor the targeted tissues are similar to the magnet detachment process. The impact of therapeutic hypothermia in the cochlea was studied by Tamames et al. \cite{b4} using a COMSOL Multiphysics model validated with experimental data. The focus of their work was on the preservation of cochlear tissues via hypothermia and the design of a cooling device. Whereas this work is an excellent example of a validated heat transfer model within the cochlea, the application and physics are different from the magnetic cochlear implant surgery problem.  

The objective of this paper is, therefore, to validate a 3D heat transfer computational model of the scala tympani, which will enable quantification of the risks of localized hyperthermia during magnetic cochlear implant surgery. In hearing research, new surgical methods or treatments are tested in phantoms \cite{b6} --\cite{b10}, cadavers \cite{b10, b11}, or animals \cite{b12} -- \cite{b26}. Here, transient temperatures measured in a 3D-printed scala tympani phantom subjected to a sudden change in its thermal environment and localized heating are used for validating the computational model. As the numerical and experimental results are in good agreement, the model is applied to determine the maximum safe input power to avoid localized hyperthermia when detaching the magnet. 

The balance of the paper is organized as follows. Section II is devoted to a description of the problem. Subsequently, the design of the experiment used to validate the heat transfer model and the associated scale analysis are discussed. Transient temperatures measured in a 3D-printed scala tympani phantom subjected to a sudden change in its thermal environment and localized heating are compared against numerical predictions in section IV. Finally, the validated heat transfer model is used to predict the maximum input power to avoid hyperthermia in the scala tympani for various boundary conditions and heating durations.

\section{Description of the problem}
\label{sec:Description of the problem}
\begin{figure}[!t]
\centerline{\includegraphics[width=\columnwidth]{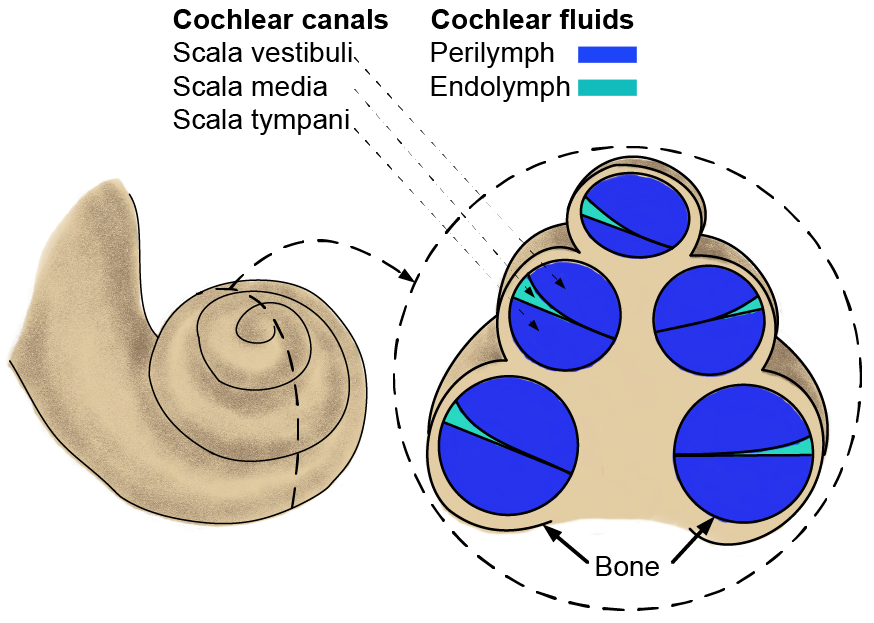}}
\caption{Cutaway view of the cochlea. The cochlea includes three canals, namely the scala tympani, the scala vestibuli, and the scala media. The cochlear implant electrode array is magnetically-guided in the scala tympani. }
\label{fig1}
\end{figure}

The inner ear consists of the cochlea and the organ of balance \cite{b23}. The cochlea is a spiral bony organ with two and a half turns consisting of three canals, namely the scala vestibuli, the scala media and the scala tympani (see Fig. \ref{fig1}) \cite{b15} -- \cite{b24}. The scala tympani and scala vestibuli are filled with a dilute saline fluid called perilymph, while the scala media is filled with endolymph which is also a dilute saline fluid \cite{b15} -- \cite{b24}. Both the perilymph and endolymph are characterized by thermophysical properties identical to that of water. The cochlear implant is magnetically guided in the scala tympani. 

\begin{figure}[!t]
\centerline{\includegraphics[width=\columnwidth]{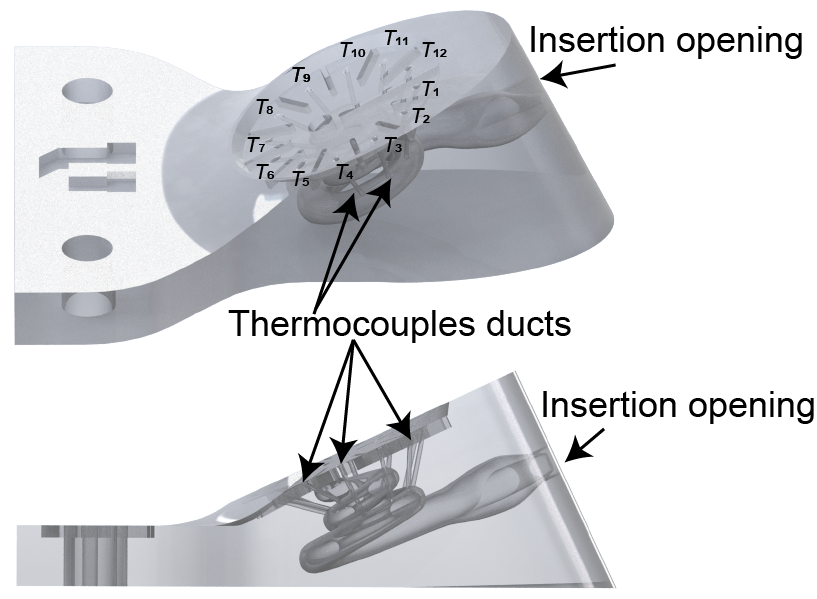}}
\caption{3D drawing of the scala tympani phantom. }
\label{fig2}
\end{figure}

Physical models of the scala tympani have been previously constructed \cite{b18, b19}. In this work, the physical model developed by Lisandro et al. \cite{b19} (see Fig. \ref{fig2}) is used for validating the computational heat transfer model within the scala tympani. To avoid confusion with the computational model, the physical scala tympani model used in the experiments will be referred to as the phantom. The phantom, consisting of a hollow scala tympani canal, was 3D-printed by Realize Inc. with a transparent material called Somos® WaterShed XC 11122. The phantom includes twelve ducts enabling the insertion of thermocouples for measuring transient temperatures along the scala tympani. A separate insertion opening included in the phantom facilitates implantation of an implant (see Fig. \ref{fig2}). Transient temperatures are measured for two distinct cases, namely when the phantom is subjected to a sudden change in its thermal environment, and when there is localized heating within the phantom. The experimental data are then compared against numerical predictions obtained with the computational heat transfer model having a geometry identical to that of the phantom. 

A deterrent to thermocouple temperature measurement at small scales is the possible influence they may have on the measurement. If the size of the thermocouples is of the same order of magnitude as the phantom, the thermocouples may affect the temperature distribution and fluid flow within the scala tympani canal, and may act as heat sinks. To avoid these potential issues, a scaled-up phantom is used in the experiments. This is discussed next. 

\section{Experimental design and scale analysis}
\label{sec:Experimental design and scale analysis}
The impact of phantom scale on the physics of the problem and the significance of each term in the energy and momentum balance equations describing fluid and heat transport is studied via a scale analysis. The scale analysis is a well-established method in thermal-fluid science that enables eliminating negligible physical phenomena in a process, and permits designing physically meaningful experiments. 

The scale analysis requires non-dimensionalizing the energy and momentum balance equations. A general form of the energy balance equation is given by \cite{b20}:

\begin{equation}
\begin{split}
\rho c_p {\frac{\partial T}{\partial t}}+{\rho c_p \mathbf{u} \cdot \nabla T}+\nabla \cdot \left(-k \nabla T \right)+\nabla \cdot {\mathbf{q}^{\prime\prime}_r}+\\
\rho_{\rm{bl}} c_{p,{\rm{bl}}} \omega_{\rm{bl}} \left(T-T_{\rm{bl}}\right) = q_{\rm{met}}+ q
\end{split}
\label{Pennes'}\end{equation}
where ${\mathbf{q}^{\prime \prime}_r}$ is the radiative heat flux vector, and $q_{\rm{met}}$ and $q$ are respectively the volumetric heat sources due to metabolic heat generation and the magnet. Blood perfusion (last term on the left-hand side of \eqref{Pennes'} in the cochlea absorbs heat and contributes to energy dissipation. The blood perfusion rate in the cochlea is less than 1 $\frac{\rm{ml}}{\rm{min}\cdot \rm{g}}$\cite{b21}. In infrared neural stimulation, the energy dissipated by perfusion in the cochlea is negligible compared to the magnitude of the heat source \cite{b2}, \cite{b3}, \cite{b4}, \cite{b5}, \cite{b21}. For example, Thompson et al. \cite{b2}, \cite{b21} discussed all possible biological cooling methods in the cochlea (perfusion, cochlear fluid flow, evaporation of water from tissues), and concluded that these mechanisms are much lower than the magnitude of the laser heat source power (6.25 mW) used for infrared neural stimulation implants. Here, an input power of 21 mW applied for 10 s is necessary to detach the magnet by assuming that the magnet is attached to the implant electrode array by 1 mm$^3$ of paraffin having a melting point of 43$^{\circ}$C. This is more than three times larger than the power reported by Thompson et al. \cite{b2}, \cite{b21}. The conclusions of Thompson et al. \cite{b2}, \cite{b21} may be applicable for magnetic cochlear implant surgery, but heat dissipation by blood perfusion should also be compared to other heat dissipation mechanisms in the scala tympani, and not solely to the external input power. 

According to \cite{b2}, the heat removal rate per unit volume by perfusion in the ear for a temperature difference of 3$^{\circ}$C and a perfusion rate of 1 $\frac{\rm{ml}}{\rm{min}\cdot \rm{g}}$ is $\sim$ 0.03 mW. For magnetic cochlear implant surgery, the temperature difference may be up to 6$^{\circ}$C, such that the perfusion heat removal rate is estimated at 0.06 mW (assuming a cochlear volume of 185.4 mm$^3$ \cite{b36} and a heat removal rate of 0.33 $\frac{\rm{kW}}{\rm{m^3}}$). In contrast, heat dissipation through the cochlear temporal bone characterized by a surface area of 250 mm$^2$ \cite{b52} and a thickness of 4 mm \cite{b51} is approximately 120 mW (assuming a cochlear volume of 185.4 mm$^3$ \cite{b36} and a heat removal rate of 647.2 $\frac{\rm{kW}}{\rm{m^3}}$ as estimated by Fourier’s law). As a result, the blood perfusion term in \eqref{Pennes'} is negligible with respect to heat dissipation by conduction through the temporal bone. 

The metabolic heat generation rate per unit volume ($q_{\rm{met}}$) in a cochlea is approximately 1.1 $\frac{\rm{kW}}{\rm{m^3}}$\cite{b4}. Under the assumption that the entire cochlea is filled with blood, which is an overestimation, the metabolic heat generation is $\sim$ 0.2 mW. This is two orders of magnitude smaller than the magnitude of the heat source (21 mW), such that the metabolic heat generation term can be neglected in \eqref{Pennes'}. 

Heat transport by radiation is estimated at 0.01 mW using the Stefan-Boltzmann law, which is three orders of magnitude smaller than the heat source. The contribution to heat transfer by radiation is therefore neglected in \eqref{Pennes'}. 

After neglecting blood perfusion, metabolic heat generation, and radiation heat transfer, the simplified energy balance equation applied to the solid magnet region where heat is transferred solely by conduction is given by: 

\begin{equation} \label{magnetenergy}
{\rho c_p \frac{\partial T}{\partial t}} + {\nabla \cdot \left(-k \nabla T\right)} = q
\end{equation}\\
The first term on the left-hand side of \eqref{magnetenergy} represents thermal energy storage, whereas the second term is the conduction heat transfer (diffusion). The energy balance in the perilymph does not include the input power density, but accounts for heat transfer by conduction and convection. The energy balance equation in the perilymph region is:
\begin{equation} \label{perilumphenergy}
{\rho c_p \frac{\partial T}{\partial t}} +{\rho c_p \mathbf{u} \cdot \nabla T} + {\nabla \cdot \left(-k \nabla T\right)} = 0
\end{equation}\\
where the second term on the left-hand side is the heat transport by convection. To calculate this term, the velocity vector should be simultaneously computed by solving the momentum balance (i.e., Navier-Stokes) equation in the perilymph: 
\begin{equation} \label{perilumphmomentum}
\rho \frac{\partial \bf{u}}{\partial t}+\rho\left(\bf{u}\cdot\nabla\right)\bf{u}=\nabla \cdot \it{p} \bf{I} + \rho\bf{g}
\end{equation}\\
where the first term on the left-hand side is the acceleration term and the second term represents advection. The first term on the right-hand side is the pressure force and the second term represents the body force (gravity) acting in the negative $z$ direction. Note that the external magnetic field guiding the magnet through the scala tympani is turned off after the surgery, such that magnetohydrodynamic effects do not come into the picture.

The momentum and energy balance equations are non-dimensionalized hereafter for the scale analysis. The dimensionless Cartesian coordinates are defined as:

\begin{equation} \label{Dimionsionlesscoordinates}
\it{X}=\frac{x}{L_c},
Y=\frac{y}{L_c},
Z=\frac{z}{L_c}
\end{equation}
where $L_c$ is a characteristic length. Similarly, the Cartesian components of the velocity vector are non-dimensionalized as follows: 

\begin{equation} \label{Dimionsionlessvelocity}
U_x=\frac{u_x}{u_0},
U_y=\frac{u_y}{u_0},
U_z=\frac{u_z}{u_0},
\end{equation}
where $u_0 (=\sqrt{g \beta L_c (T_{max}-T_i)}$) is derived by assuming that the gravitational and viscous forces are in equilibrium. The dimensionless temperature is defined as $\theta=\frac{T-T_i}{T_{max}-T_i}$, where $T_i$ is initial temperature, and the dimensionless time, also called the Fourier number, is given by $Fo=\frac{\alpha t}{L_c^2}$.

Substituting the dimensionless parameters into \eqref{magnetenergy} to \eqref{perilumphmomentum}, the dimensionless energy balance in the solid magnet region is:
\begin{equation} \label{Dimionsionlessmagnetenergybalance}
\frac{\partial \theta}{\partial Fo} = \left(\frac{\partial ^2 \theta}{\partial X^2} + \frac{\partial ^2 \theta}{\partial Y^2} +\frac{\partial ^2 \theta}{\partial Z^2}\right) + Q
\end{equation}\\
where $Q$ = $\frac{q L_c^2}{Vk(T_{max}-T_i)}$ is the dimensionless input power. The dimensionless energy balance in the perilymph is written as:
\begin{equation} \label{Dimionsionlessperilymphenergybalance}
\begin{split}
\frac{\partial \theta}{\partial Fo} + Pr Gr^{0.5} \left(U_x\frac{\partial \theta}{\partial X} + U_y\frac{\partial \theta}{\partial Y} + U_z\frac{\partial \theta}{\partial Z}\right)=\\ \frac{\partial ^2 \theta}{\partial X^2} + \frac{\partial ^2 \theta}{\partial Y^2} +\frac{\partial ^2 \theta}{\partial Z^2}
\end{split}
\end{equation}\\
where $Pr = \frac{\nu}{\alpha}$ and $Gr=\frac{g \beta (T_s - T_{\infty})L_c^3}{\nu^2}$ are the Prandtl and Grashoff numbers, respectively. Finally, the dimensionless momentum balance equations in the perilymph along each Cartesian coordinate are given by:
\begin{equation}
\label{Dimionsionlessperilymphmomentumbalancex}
\begin{split}
\frac{1}{Pr}\frac{\partial U_x}{\partial Fo} + Gr^{0.5} \left(U_x \frac{\partial U_x}{\partial X}+ U_y\frac{\partial U_x}{\partial Y} +U_z\frac{\partial U_x}{\partial Z}\right) = \\
-\left(\frac{\partial^2 U_x}{\partial X^2}+\frac{\partial^2 U_x}{\partial Y^2}+\frac{\partial^2 U_x}{\partial Z^2}\right)
\end{split}
\end{equation}

\begin{equation} \label{Dimionsionlessperilymphmomentumbalancey}
\begin{split}
\frac{1}{Pr}\frac{\partial U_y}{\partial Fo} + Gr^{0.5} \left(U_x \frac{\partial U_y}{\partial X}+ U_y\frac{\partial U_y}{\partial Y} +U_z\frac{\partial U_y}{\partial Z}\right) =\\
-\left(\frac{\partial^2 U_y}{\partial X^2}+\frac{\partial^2 U_y}{\partial Y^2}+\frac{\partial^2 U_y}{\partial Z^2}\right)
\end{split}
\end{equation}

\begin{equation}
\label{Dimionsionlessperilymphmomentumbalancez}
\begin{split}
\frac{1}{Pr}\frac{\partial U_z}{\partial Fo} + Gr^{0.5} \left(U_x \frac{\partial U_z}{\partial X}+ U_y\frac{\partial U_z}{\partial Y} +U_z\frac{\partial U_z}{\partial Z}\right) =\\
-\left(\frac{\partial^2 U_z}{\partial X^2}+\frac{\partial^2 U_z}{\partial Y^2}+\frac{\partial^2 U_z}{\partial Z^2}\right) + 
Gr^{0.5} \theta
\end{split}
\end{equation}

Based on \eqref{Dimionsionlessmagnetenergybalance} to \eqref{Dimionsionlessperilymphmomentumbalancez}, the scale analysis implies that if the dimensionless parameters $Fo, Pr, Gr,$ and $Q$ are equal for the actual physical model (prototype) and the scaled-up physical model (simply called model hereafter), then the dimensionless temperature $\theta$ and dimensionless velocities $U_x, U_y, U_z$ will be equal for the prototype and the model. 

\begin{figure}[!t]
\centerline{\includegraphics[width=\columnwidth]{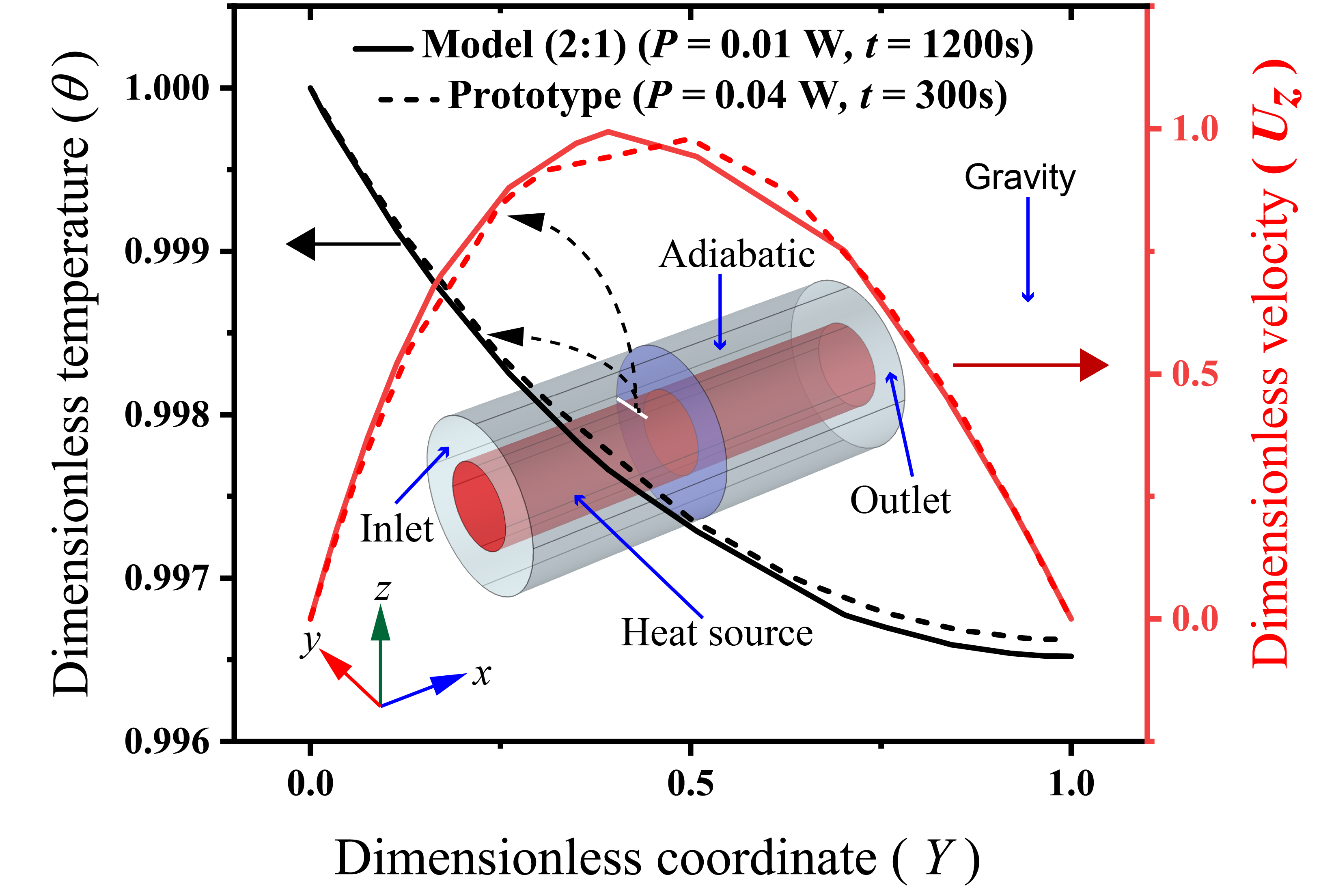}}
\caption{Dimensionless temperature and velocity as a function of the dimensionless distance from the origin for the prototype and model at ($x$=0, $y$=0, $z$=0).  }
\label{fig3}
\end{figure}

The scale analysis is first verified by solving a simple heat transfer problem between two concentric cylinders. The inner cylinder is heated via constant heat rates of 0.01 W (model) and 0.04 W (prototype). The boundary of the outer cylinder is adiabatic, and the gap region between the two cylinders is filled with water. The thermophysical properties required for the simulations are listed in Table \ref{tab1}. The initial temperature of the entire system is 20$^{\circ}$C. For the prototype, the diameter and length of the inner cylinder are 1 mm and 5 mm, respectively. The outer cylinder has a diameter of 2 mm and a length of 5 mm. Here, the model is two times larger than the prototype (see Fig. \ref{fig3}). The characteristic length is defined as the gap distance between the two cylinders. The velocity at the surface of the inner and outer cylinders is zero because of the no-slip boundary condition. Consequently, the dimensionless velocity is zero, and reaches a maximum value between the two cylinders. The dimensionless velocities for the prototype and model are in a good agreement with less than 1\% difference. The temperature at the surface of the heated inner cylinder is a maximum and decreases towards the outer cylinder. Again, differences between the dimensionless temperatures of the prototype and model are less than 1\%. The slight disparities are due to mesh differences as well as truncation and round off errors. Thus, these results demonstrate that the prototype can be scaled up without affecting the physics of the problem. 

The scale analysis also enable comparing the significance of natural convection with respect to conduction heat transfer in the scala tympani. Based on preliminary simulations in the scala tympani with an inserted implant electrode array, it is found that the first term on the left-hand side of \eqref{Dimionsionlessperilymphenergybalance} is on the order of one, the second term is on the order of -7, whereas the term on the right-hand side is on the order of four. This means that natural convection heat transfer (second term on the left-hand side of \eqref{Dimionsionlessperilymphenergybalance}) is negligible in comparison to conduction (right-hand side of \eqref{Dimionsionlessperilymphenergybalance}). This result is in agreement with our previous study performed on an uncoiled cochlea model \cite{b22}, where it was shown that natural convection has a negligible impact on the thermal management of the scala tympani during magnetic cochlear implant surgery. As such, heat transfer by natural convection is neglected in the remainder of this paper. Note that the accuracy of this assumption is later confirmed with experimental data.

\begin{figure}[!t]
\centerline{\includegraphics[width=\columnwidth]{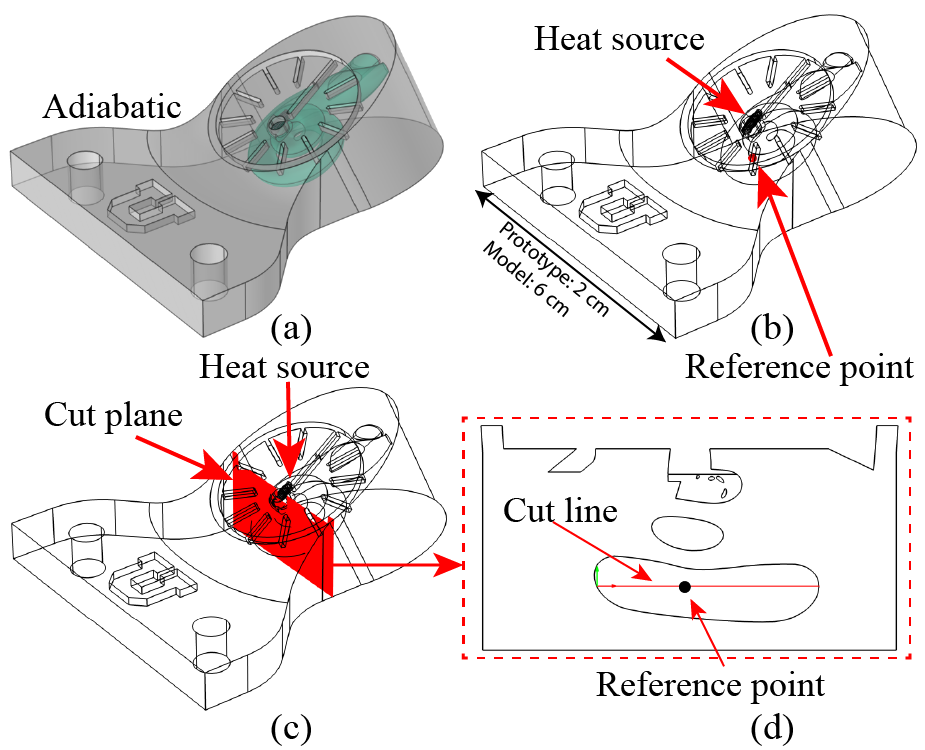}}
\caption{Scala tympani phantom: (a) Adiabatic boundary conditions are applied to all exterior walls. (b) Reference point where the transient temperature is calculated (prototype: $x$ = 0 mm, $y$ = 0 mm, $z$ = -2/3 mm, model: $x$ = 0 mm, $y$ = 0 mm, $z$ = -2 mm). (c) Cut plane crossing the reference point. (d) Cut line on the cut plane where the spatial distribution of temperature is calculated.}
\label{fig4}
\end{figure}

\begin{figure}[!t]
\centerline{\includegraphics[width=\columnwidth]{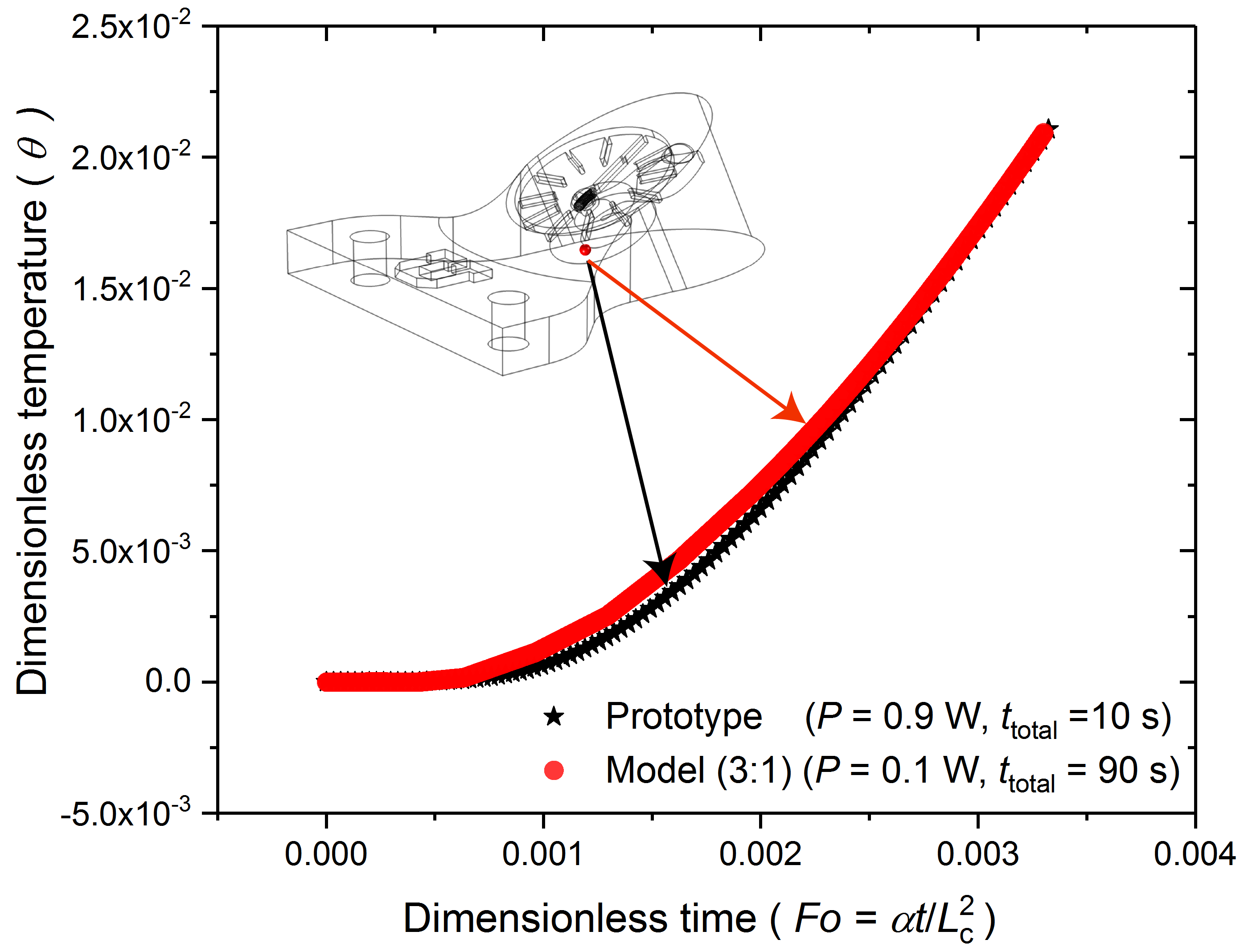}}
\caption{Dimensionless temperature as a function of the dimensionless time (Fourier number) at the reference point for the phantom prototype and model.}
\label{fig5}
\end{figure}

\begin{figure}[t!]
\centerline{\includegraphics[width=\columnwidth]{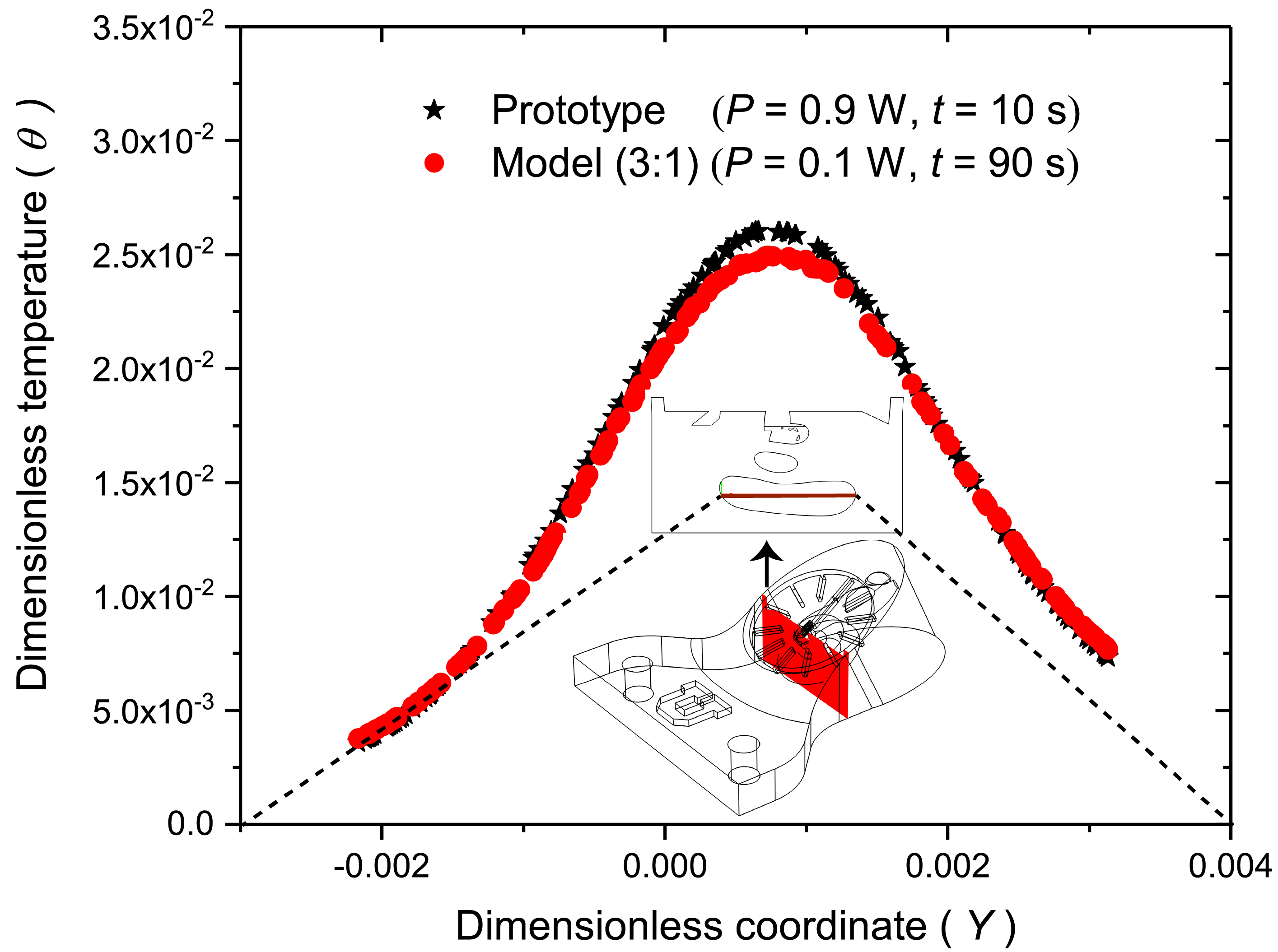}}
\caption{Dimensionless temperature as a function of the dimensionless location along the cut line defined in fig. 4(d) for the phantom prototype and model.}
\label{fig6}
\end{figure}

In addition to the two-concentric cylinder problem, the accuracy of the scale analysis for a prototype and model (3:1) of the scala tympani is also analyzed. The scale analysis is performed based on the geometry and thermophysical properties of the scala tympani phantom discussed in section II. A spiral-shaped heat source is implanted in the narrowest region of the phantom, while it is assumed that all boundaries of the phantom are adiabatic and that the initial temperature is 0$^{\circ}$C. The thermophysical properties of perilymph (same as water), the phantom prototype material (Somos® WaterShed XC11122), and the heat source (Ni-Cr80 wire) are listed in Table \ref{tab1}. The geometry of the scala tympani prototype is illustrated in Fig. \ref{fig4} (a). For equivalent values of $Q$, the power inputs for the phantom prototype and model are 0.9 W and 0.1 W, respectively. Temperatures are calculated at a reference point (prototype: $x$ = 0 mm, $y$ = 0 mm, $z$ = -2/3 mm, model: $x$ = 0 mm, $y$ = 0 mm, $z$ = -2 mm) and a cut line on the cut plane crossing the reference point (see Figs. \ref{fig4}, (b), (c) and (d)). The transient dimensionless temperature at the reference point is plotted in Fig. \ref{fig5}. As expected, the temperature increases with increasing time when the heat source is turned on. The temperature gradient increases after some time and, as a result, the temperature increase rate is enhanced. This trend is the same for both the phantom prototype and model with less than 5$\%$ difference between individual data. The differences are due to the disparate meshes for the prototype and model as well as truncation and round off errors. The dimensionless temperature along the cut line in the phantom prototype is plotted in Fig. \ref{fig6}. The temperature is maximum near the heat source. As the distance with respect to the heat source increases, the temperature gradient decreases and, consequently, the heat transfer rate drops. The spatial temperature distribution trend is identical for the phantom prototype and model with less than 5$\%$ discrepancy at a few locations. Again, the differences are due to different mesh as well as truncation and round off errors.


\begin{table}
\caption{Thermophysical properties of the substances used in the simulations. }
\label{table}
\setlength{\tabcolsep}{2.1pt}
\begin{tabular}{|l|l|l|l|}
\hline
Substance& 
Density & Thermal conductivity & Heat capacity \\
&$\rho$ [$\frac{\rm{kg}}{\rm{m^3}}$]& $k$ [$\frac{\rm{W}}{\rm{m \cdot K}}$]& $c_p$ [$\frac{\rm{J}}{\rm{kg \cdot K}}$]\\
\hline
Perilymph (Water) &992.20 \cite{b28} &0.625\cite{b28} &4176.6\cite{b28} \\
Resistive wire (Ni-Cr80) & 8400 \cite{b29} & 11.3 \cite{b29} & 450 \cite{b29} \\
Phantom material & 1125.2$^{\mathrm{a}}$  & 0.25$^{\mathrm{a}}$ & 1610$^{\mathrm{a}}$ \\
Magnet   & 430$^{\mathrm{b}}$ & 8.1$^{\mathrm{b}}$ & 7500$^{\mathrm{b}}$ \\
Electrode array & 19400$^{\mathrm{c}}$ & 28$^{\mathrm{c}}$ & 127.7$^{\mathrm{c}}$ \\
Blood & 1050 \cite{b4} & 0.52\cite{b4}& 3840\cite{b4}\\
Bone & 1908 \cite{b4} & 0.32\cite{b4}& 1313\cite{b4}\\
\hline
\multicolumn{3}{p{200pt}}{$^{\mathrm{a}}$Provided by the manufacturer (Realize Inc.)

$^{\mathrm{b}}$Provided by the manufacturer (SUPERMAGNETMAN).

$^{\mathrm{c}}$Calculated based on the information provided by MED-EL.}
\end{tabular}
\label{tab1}
\end{table}

To conclude this section, scaling up the prototype does not change the physics of the problem. Thus, experiments are performed on a phantom identical to the scaled-up model of the scala tympani (3:1) for validating the computational heat transfer model. This is discussed in the next section. 

\section{Validation of the computational heat transfer model with experimental data}
\label{sec:Validation of the computational heat transfer model with experimental data}

\begin{figure}[!t]
\centerline{\includegraphics[width=\columnwidth]{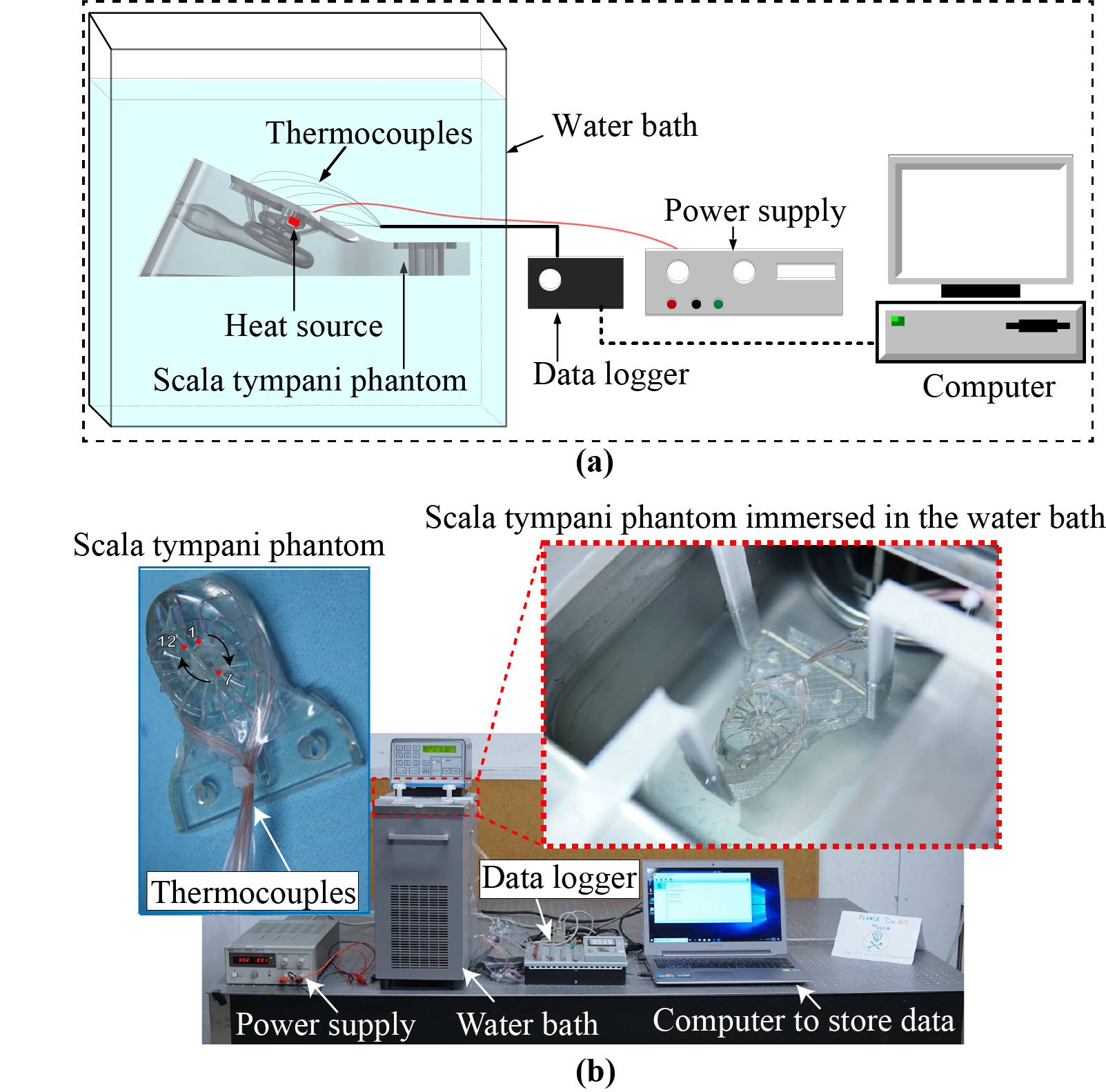}}
\caption{Schematic (panel (a)) and photographs (panel (b)) of the experimental setup. }
\label{fig7}
\end{figure}

The validation experiments are performed using a phantom model of the scala tympani with a scale of 3:1. The details of the phantom geometry are provided in \cite{b19}, and the phantom STL files are available in \cite{b56}. The phantom is equipped with 12 thermocouples (CHAL-015-BW-Omega) to measure the transient temperature along the scala tympani. A data logger (CAMPBELL SCIENTIFIC INC. CR5000) is used to record the temperature data. A water bath (VWR 1167 Heated Refrigerated Circulating Water Bath Polyscience) provides a constant and uniform temperature at the phantom boundaries. The heat source is made of a resistance wire (Omega-NIC80-010-062) connected to two copper wires (Copper Wire 30AWG) and is controlled via a power supply (E3616A 60W Power Supply, 35V, 1.7A). All instruments are calibrated prior to each experiment based on their respective user manuals. A schematic and photographs of the experimental setup are shown in Fig. \ref{fig7}.

Experimental data are used for validating the computational heat transfer model. For that purpose, the 3D geometry of the phantom has been imported in COMSOL Multiphysics. The thermocouples and their associated insertion holes have been removed in the numerical model to reduce the computational load. This simplification does not affect the results in a perceptible manner owing to the small size of the thermocouples. Based on simulations, it was mentioned in section III that natural convection has a negligible impact on the thermal management of the cochlea. This is explored further in this section by comparing the experimental data against the numerical results generated by considering only conduction within the scala tympani. 

Preliminary simulations revealed that 5,178,756 mesh elements and a time step of 0.1 s were sufficient to obtain converged results. Two scenarios are tested hereafter, namely heat transfer within the scala tympani phantom subjected to a sudden change in its thermal environment, and heat transfer within the scala tympani phantom subjected to localized heating via a small heat source.

\subsection{Scala-tympani subjected to a sudden change in its thermal environment}
\label{sec:Scala-tympani subjected to a sudden change in its thermal environment}
In these experiments, the phantom is filled with distilled water and its two outlets are plugged with rubber. The phantom is placed in a mixture of ice and water to reach a steady-state temperature corresponding to the melting point of water in the Salt Lake City geographical area, which is $\sim$ 0.004$^{\circ}$C. Meanwhile, the water bath is heated to 25$^{\circ}$C. Once steady-state temperatures are reached, the phantom is subjected to a sudden change in its thermal environment by quickly transferring it to the water bath until it reaches a temperature of 25$^{\circ}$C. The transient temperature of each thermocouple is recorded during this process. 

For the numerical simulations, the entire system is assumed to be at an initial temperature of 0$^{\circ}$C. A convective boundary condition at the phantom boundary with a constant ambient temperature of 25$^{\circ}$C is modeled. A heat transfer coefficient $h$ = 1000 $\rm{\frac{W}{m^2 \cdot K}}$ is determined by fitting experimental data with numerical simulations. This value falls within the range of possible heat transfer coefficients for forced convection in the water bath, which is between $\sim$500 and 3,000 $\rm{\frac{W}{m^2 \cdot K}}$\cite{b54,b55}. Note that natural convection in the scala tympani is neglected.

\begin{figure}[!t]
\centerline{\includegraphics[width=\columnwidth]{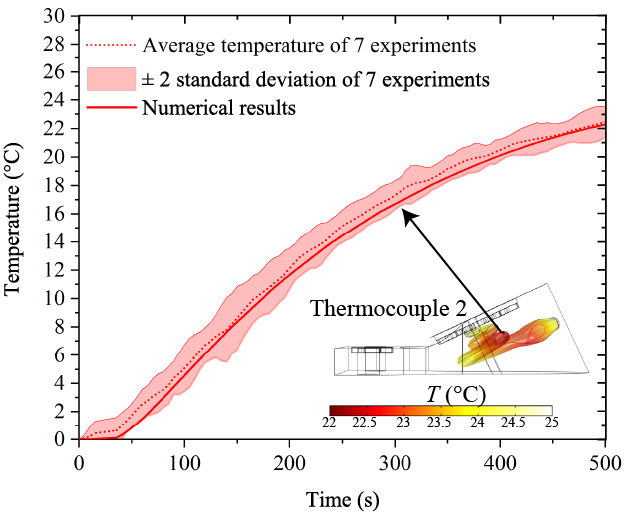}}
\caption{ Temperature as a function of time when the scala tympani is subjected to a sudden change in its thermal environment. Experimental data are compared against numerical simulations.}
\label{fig8}
\end{figure}

A sample validation result for thermocouple 2 is plotted in Fig. \ref{fig8}. The rate of temperature change is almost zero for $t < 50$ s because heat has not penetrated far enough into the phantom (i.e., thermocouple 2 does not yet “feel” the boundary condition). After 50 s, the heat from the water bath penetrates far enough in the phantom to increase the temperature at the thermocouple 2 location. The temperature increase continues at a lower rate after 350 s and approaches 22.4$^{\circ}$C at 500 s. Due to the low heat transfer rate, it takes longer than 500 s for all points within the phantom to reach thermal equilibrium with the water at 25$^{\circ}$C.

The validation plots are similar for the other 11 thermocouples. An error analysis has been conducted based on the ASME V$\&$V 20 2009 Standard for Verification and Validation in Computational Fluid Dynamics and Heat Transfer \cite{b30}. To quantify the simulation error, the absolute difference between the average temperature recorded in seven experiments and the calculated temperature for each instant is calculated. Then, the maximum difference between the numerical and experimental results for each thermocouple is divided by the maximum temperature difference, which is 25$^{\circ}$C. Using this technique, it is found that the average error for all 12 thermocouples is less than 5$\%$. This difference is due to numerical errors, and the uncertainties in the input properties and the initial temperature. In addition, the short time required for transferring the phantom from the ice bath to the warm water bath is not taken into account in the numerical model, which may cause minor errors. Here, given the good agreement between experimental and numerical results, it is concluded that the model supports the assumption that natural convection within the scala tympani is negligible.

\subsection{Scala tympani subjected to localized heating}
\label{sec:Scala tympani subjected to localized heating}
\begin{figure}[!t]
\centerline{\includegraphics[width=\columnwidth]{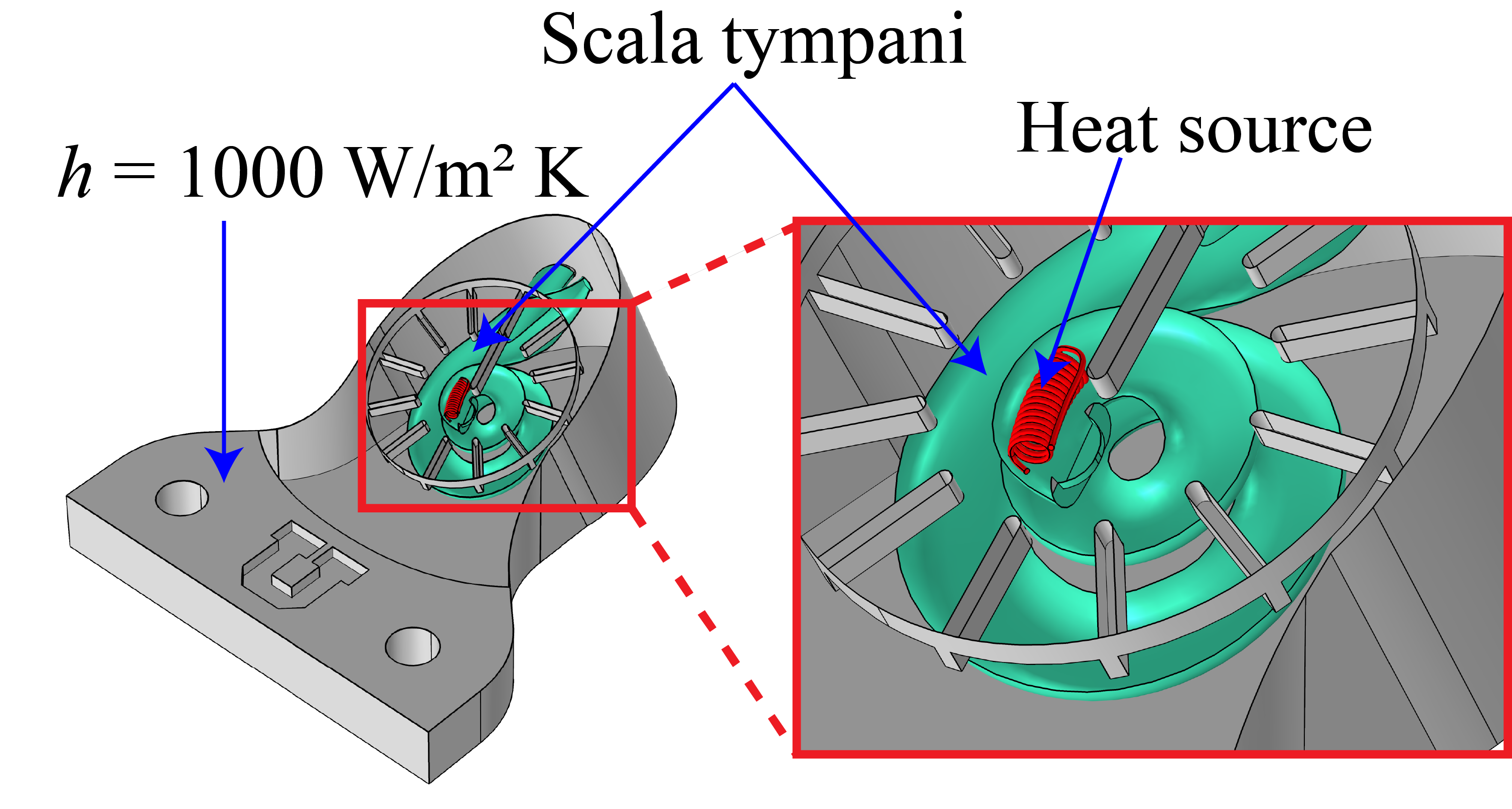}}
\caption{3D model of the scala tympani subjected to localized heating with a heat source used in the simulations for comparison against experimental data. The top surface of the phantom is removed for better visualization of the scala tympani.}
\label{fig9}
\end{figure}

Localized heating is provided by a heat source connected to the power supply and embedded close to the apex of the scala tympani (i.e., narrowest part of the scala tympani at the top of the phantom) as shown in Fig. \ref{fig9}. The scala tympani canal is filled with distilled water and is immersed in the water bath. The scala tympani canal is plugged to avoid mixing the water inside the bath with the distilled water within the phantom. The temperature of the water bath is set to 11$^{\circ}$C, and the temperature readings are monitored until all thermocouples show steady-state temperatures equal to 11$^{\circ}$C. The voltage and current supplied to the heat source are controlled by the power supply. As the heater circuit is in series, the input current is the same in the whole circuit. The total resistance of the circuit is slightly higher than the resistance of the resistive wire (Ni-Cr80) located inside the phantom. The total resistance of the circuit includes the resistance of the copper wires in addition to the resistance of the Ni-Cr80 wire. The resistance of the Ni-Cr80 wire ($R$ = 2.1 $\Omega$) and the input current ($I$ = 0.3 A) are used to calculate the input power. Transient temperatures of all 12 thermocouples are recorded for validating the computational heat transfer model. For the simulations, the initial temperature of the entire system is set to 11$^{\circ}$C. A convective boundary condition with an ambient temperature of 11$^{\circ}$C and a heat transfer coefficient of $h$ = 1000 $\rm{\frac{W}{m^2 \cdot K}}$, as determined in the experiments without a heat source, is used. The heat source provides a power of 189 mW. 

\begin{figure}[!t]
\centerline{\includegraphics[width=\columnwidth]{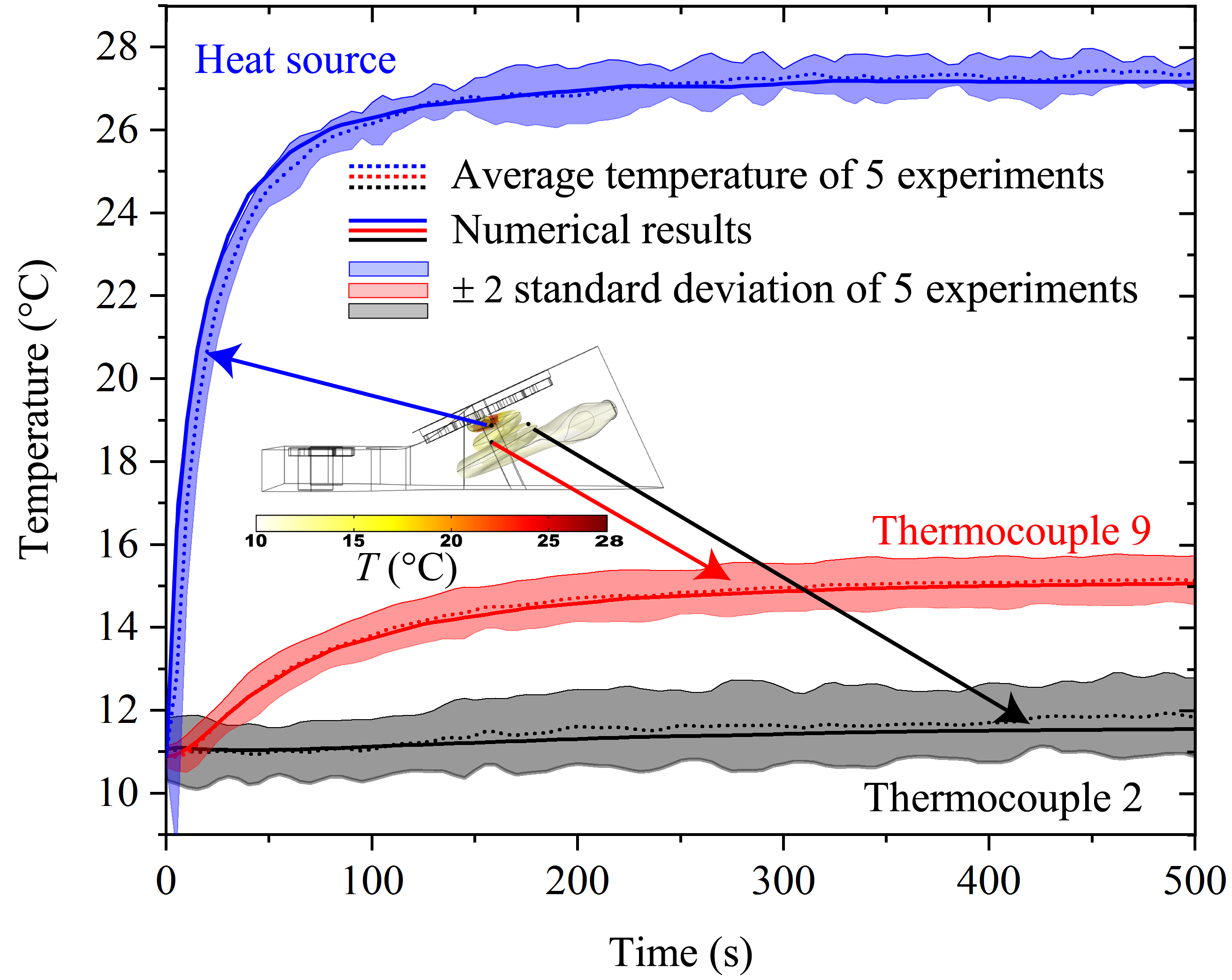}}
\caption{Temperature as a function of time when the scala tympani is subjected to localized heating. Experimental data are compared against numerical simulations. }
\label{fig10}
\end{figure}

It is found that the average difference between the model and experiment data, considering all 12 thermocouples, is less than 6$\%$ by using the technique described in section IV.A and repeating each experiment five times. The error is due to the uncertainties in the input data, including the heat transfer coefficient and the magnitude of the heat source input power, in addition to the previously mentioned numerical errors. Fig. \ref{fig10} shows sample validation results for thermocouples 2, 9, and the thermocouple embedded within the heat source. Similar trends are obtained for all thermocouples. The temperature of the heat source increases within 200 s to $\sim$27$^{\circ}$C, after which steady-state conditions are reached where the heat removal rate balances the input power. As expected, the rate of temperature change at other locations within the scala tympani decreases with increasing distance from the heat source. For example, the temperature recorded by thermocouple 2 (located 9.6 mm away from heat source) changes minimally during the transient process, while the temperature measured by thermocouple 9 (located 6.8 mm away from the heat source) increases from 11$^{\circ}$C to 15$^{\circ}$C.


In conclusion, a good agreement between experimental data and numerical results has also been obtained with localized heating. As such, the 3D heat transfer model of the scala tympani can be applied with confidence to predict the maximum safe input power to avoid localized hyperthermia when detaching the magnet after magnetic cochlear implant surgery.


\section{Maximum safe input power to avoid localized hyperthermia}
\label{sec:Maximum safe input power to avoid localized hyperthermia}

The validated 3D heat transfer model of the scala tympani is applied hereafter to calculate the maximum safe input power to avoid localized hyperthermia when detaching the magnet from the implant electrode array. The maximum safe input power is determined using the thermal dose or cumulative equivalent minutes (CEM). The CEM evaluates thermal damage of tissues as a function of temperature and length of exposure to an elevated temperature. The thermal threshold of different types of tissues are in the range of 40$^{\circ}$C  to 55$^{\circ}$C \cite{b32, b33}. The death rate of most types of tissues increases at 43$^{\circ}$C, such that this temperature is selected as a reference for calculating the CEM. The safe length of exposure to 43$^{\circ}$C for coclear tissues is 114 s \cite{b34}. This means that cochlear tissues are not damaged if they are exposed to a temperature of 43$^{\circ}$C during 114 s after which the temperature suddenly drops to the body core temperature. During the actual process, tissues are heated from the body core temperature to 43$^{\circ}$C, or tissues may even be heated to temperatures higher than 43$^{\circ}$C. After turning off the heat source, cochlear tissues cool down gradually, and, as a result, they stay at a temperature higher than the body core temperature for a certain amount of time after detaching the magnet. This cooling period is taken into account in the calculation of thermal dose which is given by \cite{b32, b33}:

\begin{equation} \label{thermaldose}
\rm{{CEM}_{43}}=\sum_{\it{t}=\rm{0}}^{\it{t}=\it{t}_i} \Delta \it{t}_n \it{B}^{{(\rm{43}-\it{T}_{n})}}
\end{equation}
where $t_i$ is the total length of exposure to all temperatures, $\Delta t_n$ is the length of exposure to the specific temperature $T_n$, and $B$ is a constant equal to 0.25 for temperatures lower than 43$^{\circ}$C and 0.5 for temperatures higher than 43$^{\circ}$C. In the simulations, the magnet is heated at discrete input power densities, for a fixed period of time, then the heating process is stopped and the simulation is continued until the maximum temperature in the cochlea, implant electrode array and magnet is less than 37.05$^{\circ}$C. The $\rm{CEM}_{43}$ for the whole process, including heating and cooling, is calculated. The maximum power density providing a $\rm{CEM}_{43}$ smaller than 114 s but larger than 113 s is selected as the maximum safe input power density to detach the magnet. 

\begin{figure}[!t]
\centerline{\includegraphics[width=\columnwidth]{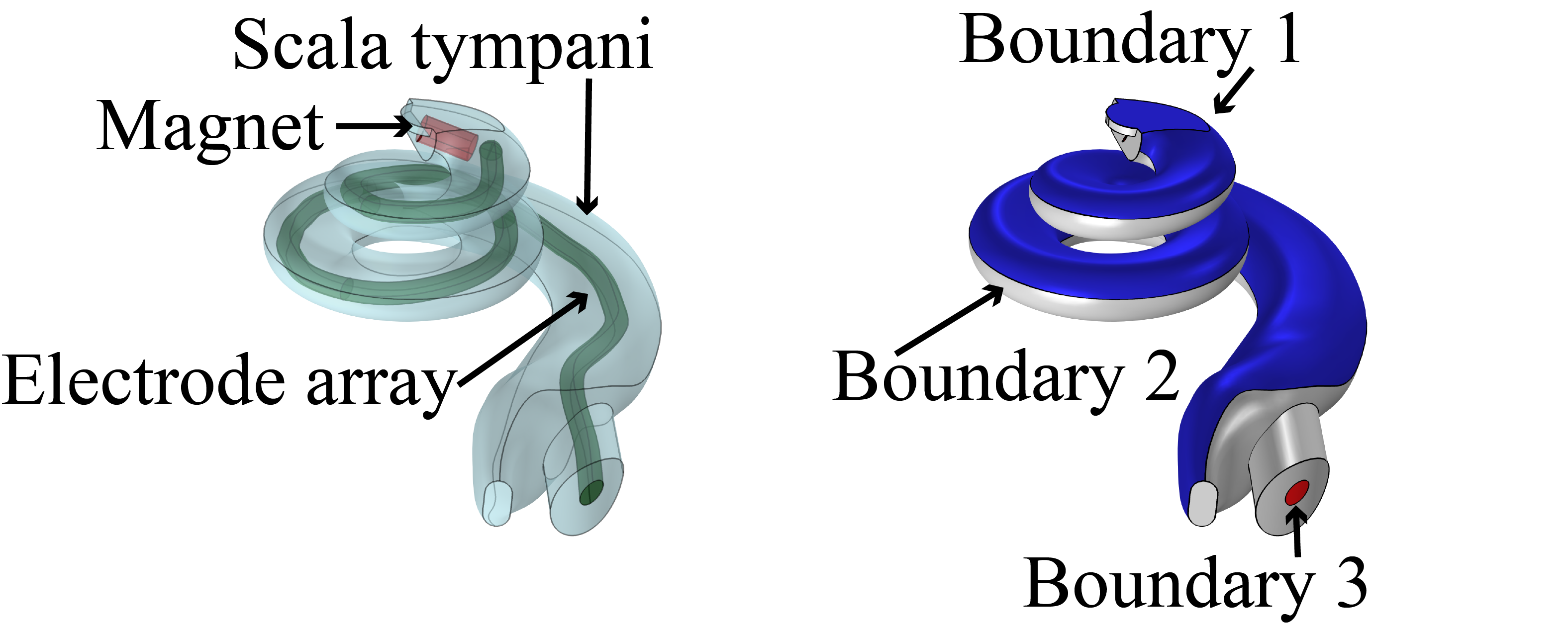}}
\caption{3D model of the scala tympani with inserted magnet and cochlear implant electrode array.}
\label{fig11}
\end{figure}

The geometry of the scala tympani numerical model used in section IV is modified for calculating the maximum safe input power. Simulations are performed in the previous section on a model involving a hollow scala tympani within a surrounding material in order to reproduce as faithfully as possible the 3D-printed phantom. Modeling a finite surrounding material is not required when dealing with the actual situation where the implant electrode array is inserted in the scala tympani. Therefore, to find the maximum safe input power to detach the magnet, the finite material surrounding the scala tympani and the resistive wire that was used as the heat source are removed from the model used in section IV. Instead, the implant electrode array and the magnet are added to the model (see Fig. \ref{fig11}). The magnet is modeled as a cylinder with a diameter of 0.5 mm and length of 1 mm. The electrode array is modeled as a 31-mm-long, 0.2-mm-diameter cylinder. The thermophysical properties of the implant electrode array and the magnet are listed in Table \ref{tab1}. In Fig. \ref{fig11}, the membrane between the scala tympani and scala media is defined as boundary 1, while boundaries 2 and 3 are respectively the bony boundary of the scala tympani and the insertion opening. The scala media has a relatively small thickness in comparison to the scala vestibuli and scala tympani, such that it is assumed that the scala media and scala vestibuli form a single region. Boundary 1 is assumed to be either isothermal at the body core temperature of 37$^{\circ}$C or adiabatic, boundary 2 is assumed to be isothermal, adiabatic, or subjected to convection, while boundary 3 is always isothermal at the body core temperature. The five different sets of boundary conditions considered in the simulations are listed in Table \ref{tab2}. 

\begin{figure}[!t]
\centerline{\includegraphics[width=\columnwidth]{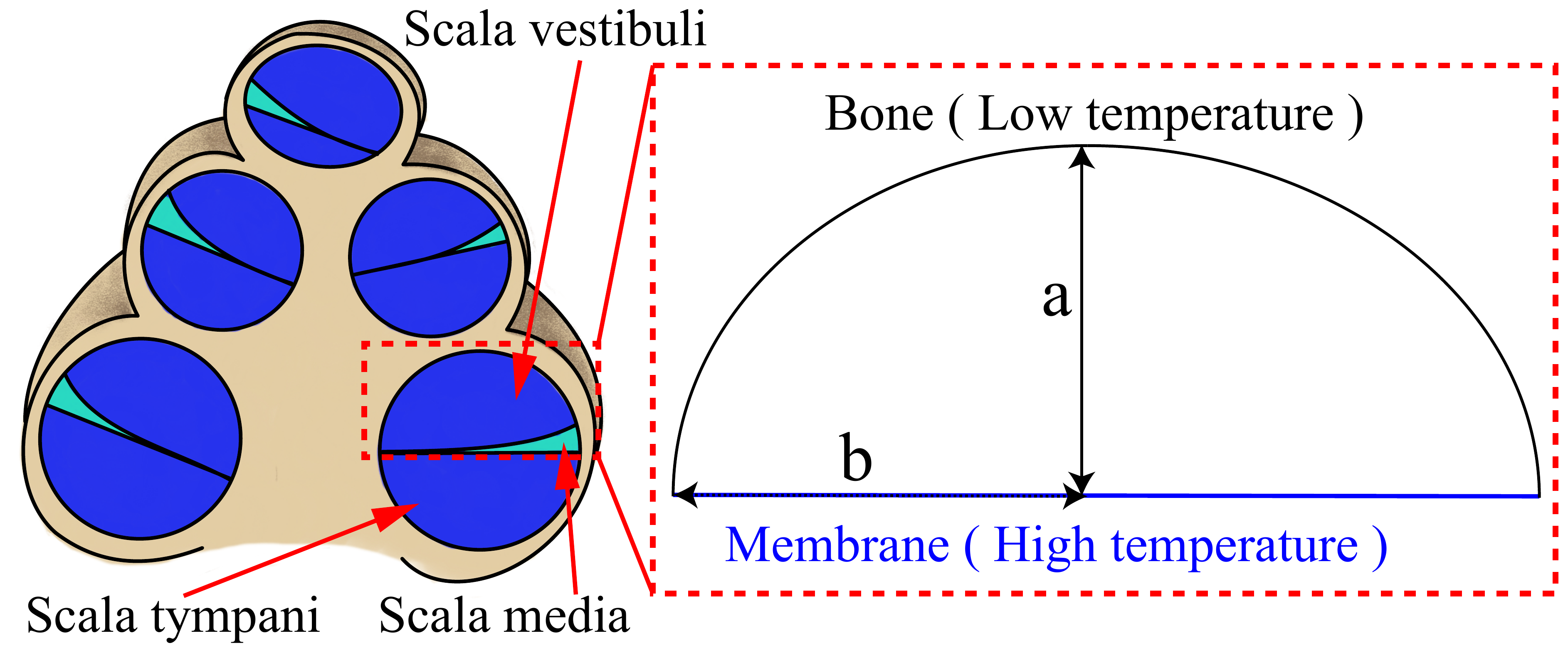}}
\caption{Characteristic length used to estimate the heat transfer coefficient from the scala tympani to the scala vestibuli and scala media.}
\label{fig12}
\end{figure}

It is assumed that heat is transferred from the scala tympani to the scala media and scala vestibuli by convection. The possible values for the heat transfer coefficient at boundary 1 are estimated by evaluating the magnitude of the Nusselt number $Nu$. The scala media and scala vestibuli are described by a semi-elliptical cross-section (see Fig \ref{fig12}), and it is assumed that the membrane is at 43$^{\circ}$C and the bony wall is at 37$^{\circ}$C. The Rayleigh number $Ra$ is less than 1000 within the scala tympani, scala media, and scala vestibuli based on a 6°C temperature difference, and a characteristic length of 1.64 mm. Bouras et al. \cite{b31} calculated $Nu$ for a semi-elliptical geometry and showed that for $Ra$ less than 5000, $Nu$ is equal to 1. The Nusselt number $Nu$ is defined as $\frac{h\cdot L_{c}}{k_{fluid}}$, thus $h=\frac{Nu\cdot k_{fluid}}{L_{c}}$. The characteristic length is usually defined as the thickness of the boundary layer. In this study, the length $a$ as defined in Fig. \ref{fig12} is assumed to be the characteristic length. The diameter of the scala media is negligible in comparison to the diameter of the scala vestibuli. The diameter of the scala vestibuli varies from $\sim$1.64 mm at the base of the cochlea to $\sim$0.81 mm at the apex of the cochlea \cite{b35}. As such, this diameter range is used to estimate the heat transfer coefficient. The minimum and maximum heat transfer coefficients are 385 $\rm \frac{W}{m^2\cdot K}$ and 775 $\rm \frac{W}{m^2\cdot K}$. These two extreme values are tested in scenarios 4 and 5 (see Table \ref{tab2}). 

\begin{figure}[!t]
\centerline{\includegraphics[width=\columnwidth]{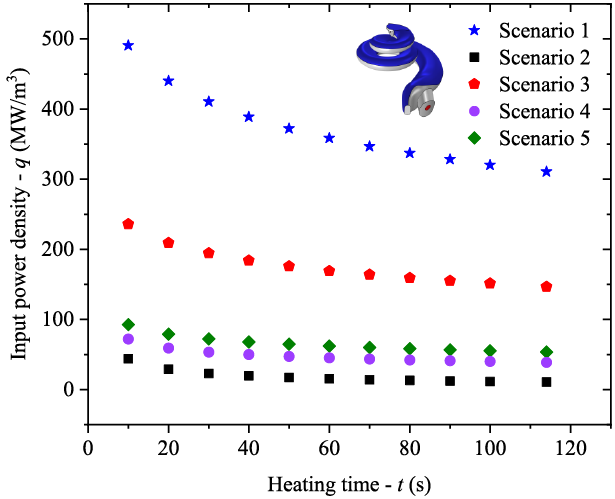}}
\caption{Maximum safe input power density to avoid hyperthermia when detaching the magnet as a function of the heating time and sets of boundary conditions defined in Table \ref{tab2}. }
\label{fig13}
\end{figure}

The maximum safe input power density to avoid localized hyperthermia when detaching the magnet is calculated for eleven discrete heating periods and for the five sets of boundary conditions listed in Table \ref{tab2}. The results are plotted in Fig. \ref{fig13}. As expected, the maximum safe input power density increases by reducing the heating period. Among the five sets of boundary conditions considered, scenario 2 where boundaries 1 and 2 are adiabatic is the most conservative. For a given heating period, scenarios 4 and 5 involving convection at boundary 1 are the closest to the adiabatic boundary condition. Therefore, the adiabatic boundary condition (scenario 2) has the highest safety factor and ensures that tissues will not suffer thermal damage during the magnet detachment process. 

As discussed in section \ref{sec:Experimental design and scale analysis}, assuming that the magnet is attached to the implant electrode array by 1 mm$^3$ of paraffin, a power of 21 mW during 10 s is necessary to melt the paraffin at 43$^{\circ}$C ($\sim$21 $\times$ 10$^{6}$~{W/m$^{3}$}). These calculations assume that the paraffin boundaries are adiabatic and that all the heat is transferred from the magnet to the paraffin. The maximum calculated safe input power when boundaries 1 and 2 are assumed to be adiabatic is 43.9 mW during 10 s~($\sim$43.9 $\times$ 10$^{6}$~{W/m$^{3}$}), which is higher than the energy required to melt the paraffin. 

To conclude, the magnet can be detached from the implant electrode array without causing localized hyperthermia. For the most efficient and safest result, it is better to apply higher input power density in a shorter period of time and use a grade of paraffin that has a melting temperature lower than 43$^{\circ}$C but higher than 38$^{\circ}$C.

\begin{table}
\fontsize{5}{5}
\caption{Sets of boundary conditions used to calculate the maximum safe input power density. }
\notsotiny
\label{table}
\setlength{\tabcolsep}{1pt}
\begin{tabular}{|l|l|l|l|}
\hline
&
Boundary 1& 
Boundary 2& 
Boundary 3\\
\hline
Scenario 1 & Isothermal (37$^{\circ}$C) & Isothermal (37$^{\circ}$C) & Isothermal (37$^{\circ}$C) \\
Scenario 2 & Adiabatic & Adiabatic & Isothermal (37$^{\circ}$C)\\
Scenario 3 & Isothermal (37$^{\circ}$C) & Adiabatic & Isothermal (37$^{\circ}$C)  \\
Scenario 4 & Convection ($h = 385 \rm{\frac{W}{m^2 \cdot K}}, \it{T_{\infty}}$=37$^{\circ}C$) & Adiabatic & Isothermal (37$^{\circ}$C) \\
Scenario 5   & Convection ($h = 775 \rm{\frac{W}{m^2 \cdot K}}$, $\it{T_{\infty}}$=37$^{\circ}C$) & Adiabatic & Isothermal (37$^{\circ}$C) \\
\hline
\end{tabular}
\label{tab2}
\end{table}

\section{Conclusions}
\label{sec:Conclusions}
{A 3D heat transfer model of the scala tympani has been validated using a phantom subjected to a sudden change in its thermal environment and to localized heating. Comparison of measured and predicted transient temperatures revealed that the average errors were less than 5$\%$ and $6\%$ without and with a heat source, respectively. This work also confirmed that natural convection heat transfer has a negligible impact on the thermal management of the scala tympani with respect to conduction heat transfer. Finally, the validated heat transfer model was applied to determine the maximum safe input power to avoid localized hyperthermia when detaching the magnet from the implant electrode array. Five different combinations of boundary conditions and eleven discrete heating periods were considered in the simulations. It was shown that the power required to release the magnet attached to the implant electrode array by 1 mm$^3$ of paraffin was less than the maximum safe input power calculated for the most conservative case where all boundaries, except the insertion opening, are adiabatic. The results of this study will enable designing a thermally safe procedure for magnetic cochlear implant surgery. In addition, the validated 3D heat transfer model is not limited to cochlear implant surgery and can be applied to other thermal management problems related to hearing science. }

\section*{Acknowledgments}
Research reported in this publication was supported by the National Institute on Deafness and Other Communication Disorders of the National Institutes of Health under Award Number R01DC013168. The content is solely the responsibility of the authors and does not necessarily represent the official views of the National Institutes of Health. We acknowledge the Center for High Performance Computing at the University of Utah for their support and resources, and in particular Dr.\ Martin Cuma for his assistance in facilitating the numerical computations. We would also like to show our gratitude to the MED-EL company for sharing the information about their electrode arrays. We would like to thank Dr. Lisandro Leon and Matt Cavilla for designing and fabricating the scala tympani phantom and Travis Morrison for helping to set up the experiments.

\end{document}